# Observation of the Dipole Blockade Effect in Detecting Rydberg Atoms by the Selective Field Ionization Method

E. A. Yakshina [a,b], D. B. Tretyakov [a,b], V. M. Entin [a,b], I. I. Beterov [a,b,c], I. I. Ryabtsev [a,b,*]

[a] *Rzhanov Institute of Semiconductor Physics, SB RAS, pr. Lavrentyeva 13, 630090 Novosibirsk, Russia*

[b] *Novosibirsk State University, ul. Pirogova 2, 630090 Novosibirsk, Russia*

[c] *Novosibirsk State Technical University, pr. Marksa 20, 630073 Novosibirsk, Russia*

\* e-mail: ryabtsev@isp.nsc.ru

**Abstract -** The dipole blockade effect at laser excitation of mesoscopic ensembles of Rydberg atoms lies in the fact that the excitation of one atom to a Rydberg state blocks the excitation of other atoms due to the shift in the collective energy levels of interacting Rydberg atoms. It is used to obtain the entangled qubit states based on single neutral atoms in optical traps. In this paper, we present our experimental results on the observation of the dipole blockade for mesoscopic ensembles of 1–5 atoms when they are detected by the selective field ionization method. We investigated the spectra of the three-photon laser excitation $5S_{1/2} \to 5P_{3/2} \to 6S_{1/2} \to nP_{3/2}$ of cold Rb Rydberg atoms in a magneto-optical trap. We have found that for mesoscopic ensembles this method allows only a partial dipole blockage to be observed. This is most likely related to the presence of parasitic electric fields reducing the interaction energy of Rydberg atoms, the decrease in the probability of detecting high states, and the strong angular dependence of the interaction energy of Rydberg atoms in a single interaction volume.

## 1. INTRODUCTION

Atoms in highly-excited Rydberg states with a principal quantum number $n \gg 1$ are the subject of intensive studies at present. Since the orbit radius of a Rydberg electron grows as $n^2$ with increasing $n$, the dipole moments of Rydberg atoms also grow as $n^2$, and they interact with one another much more strongly than do atoms in the ground state [1]. This property of Rydberg atoms is used to realize quantum computers and simulators with qubits based on single neutral atoms of alkali metals in optical dipole traps [2–4]. Short-term laser excitation of atoms to Rydberg states allows the interactions between qubits to be switched on and off, which is needed to perform quantum gates or quantum simulations [5].

One of the main methods for performing quantum gates is to use the change in the collective excitation probability of an ensemble of interacting Rydberg atoms (dipole blockade effect). Its essence is that in the presence of interaction the excitation of one Rydberg atom in a small volume shifts the resonance frequencies and blocks the excitation of other atoms; therefore, only one Rydberg atom can be excited from the entire mesoscopic ensemble. The dipole blockade was first predicted in [6] and then observed in various experiments. It was used to obtain the entangled qubit states based on single neutral atoms in optical traps [7].

Large ensembles $N=10^3$-$10^5$ of cold Rb or Cs atoms in magneto-optical traps were used in the first experiments to demonstrate the signatures of dipole blockage [8–13]. Rydberg atoms were detected by the selective field ionization (SFI) method with the detection of ions by electron multipliers based on microchannel plates. In this method atoms are ionized with a probability of 1 as soon as the electric field reaches a critical value

$$E_{cr} \approx 3.2 \times 10^8 / n_{eff}^4 \text{ V/cm},  \qquad (1)$$

where $n_{eff} = n - \delta_L$, and $\delta_L$ is the quantum defect of the Rydberg state that depends on its orbital angular moment $L$ [1]. The SFI method is characterized by a high speed (microseconds) and a high detection efficiency (more than 50%). However, no complete dipole blockade, when only one atom is excited from the entire large ensemble, was observed in such experiments. Instead, a decrease in the excitation probability of high Rydberg states ($n \geq 80$) by 30–50%, which should be described by the scaling dependence $n_{eff}^{-3}$ for noninteracting atoms, was recorded. No experiments with small mesoscopic ensembles ($N$=1-10 атомов) were performed, because the microchannel plate multipliers were used in the analog regime and had no resolution in the number of atoms.

The succeeding experiments to observe the dipole blockade were aimed at implementing two-qubit gates and, therefore, were conducted with two atoms in neighboring microscopic optical dipole traps 1–3 μm in diameter and with a distance between them of 5–10 μm [14, 15]. Since the individual detection of each of the atoms was required, the SFI method in such experiments was inapplicable, because all Rydberg atoms are ionized in it if a uniform electric field is applied. Instead, the much slower optical method was used. In this method single Rydberg atoms were detected by the fluorescence of atoms in the first excited state when they were illuminated by laser radiation resonant to the transition from the ground state. If an atom did not fluoresce, then this implied that it was in a Rydberg state. To detect the fluorescence of single atoms, it was necessary to use highly sensitive photon counters based on avalanche photodiodes or low-noise cooled CCD cameras with amplifiers; in this case, the minimum detection time was no less than 1 ms, which is longer than the SFI detection time by three orders of magnitude.

Subsequently, both SFI [16, 17] and optical [18–24] methods were applied to study the dipole blockade. The experiments to observe sub-Poissonian statistics [10, 16] and spatial correlations [17] under laser excitation of large ensembles of Rydberg atoms under dipole blockade conditions were performed using the SFI method. Single-atom collective excitations due to the dipole blockade not only for two atoms in optical traps [18–22], but also for large ensembles of hundreds of atoms in an optical trap or lattice [23, 24] were observed using the optical method.

Whether a complete dipole blockade can be observed by the SFI method in small mesoscopic ensembles of atoms still remains an open question. Previously we have developed an original technique for detecting such ensembles of Rydberg atoms ($N$ = 1–5) by the SFI method when using a VEU-6 channel electron multiplier [25]. It was found that the channel multiplier allowed one to distinguish the number of detected atoms and to sort the measured signals by the number of atoms after each laser pulse. Based on this technique, we performed a number of experiments to observe the electrically controlled resonant dipole-dipole interaction in mesoscopic ensembles of Rydberg atoms [26–29].

By now we have created a narrow-band laser excitation system to realize the dipole blockade and demonstrated the possibility of obtaining narrow (less than 2 MHz in width) three-photon multiatom resonances for the Rydberg state $37P_{3/2}$ [30]. This state has a weak interaction (less than 1 MHz at a distance between atoms of ~10 μm) and is unsuitable for observing the dipole blockade. Therefore, in [31] we performed a theoretical analysis of the multiatom excitation signals and showed that the dipole blockade could be observed in principle for high Rydberg states with $n \approx 120$, which should manifest itself as a change in both the amplitudes of multiatom three-photon resonances and the detection statistics of multiatom signals by the VEU-6 multiplier.

In this paper we present our experimental results on the observation of the dipole blockade for mesoscopic ensembles of $N = 1–5$ atoms in a single trap when they are detected by the SFI method. We investigated the multiatom spectra of the three-photon laser excitation $5S_{1/2} \to 5P_{3/2} \to 6S_{1/2} \to nP_{3/2}$ of cold Rb Rydberg atoms localized in a small excitation volume (~20 μm in size) in a magneto-optical trap. The signal post-selection by the number of detected atoms $N = 1–5$ was made using an original technique. It was expected that only one atom could be excited to a Rydberg state from the entire mesoscopic ensemble under a complete dipole blockade. Therefore, the dipole blockade should lead to a radical change in the multiatom spectra: the amplitude of the single-atom spectrum with $N = 1$ should increase, while all the remaining multiatom resonances should disappear. If they do not disappear completely, then this may suggest an incomplete dipole blockade, while a change in the ratio of the multiatom resonance amplitudes should allow the degree of completeness of the dipole blockade to be determined under specific experimental conditions. We also investigated the change in the detection statistics of multiatom signals by the VEU-6 multiplier under the dipole blockade.

## 2. DIPOLE BLOCKADE EFFECT AT LASER EXCITATION OF MESOSCOPIC ENSEMBLES OF RYDBERG ATOMS

The dipole blockade effect at the excitation of a mesoscopic ensemble of neutral atoms to Rydberg states by narrow-band continuous-wave laser radiation is as follows. After the excitation of one Rydberg atom by a laser pulse, the long-range interactions shift the Rydberg level degenerate with it in energy in the neighboring atoms by $\delta W$ dependent on the interaction energy. If the laser excitation line width $\delta \nu$ is much less than $\delta W$, then after the excitation of one atom to a Rydberg state, it will be impossible to excite the remaining atoms interacting with it [6].

It is most convenient to consider the interaction of atoms in Rydberg states under the conditions of Förster resonances, which we investigated previously [26–29]. Such resonances emerge, for example, when the Rydberg level 2 excited by laser radiation from the ground state 0 lies midway between the two adjacent levels 1 and 3 of opposite parity (Fig. 1a). The Förster resonances $nP_{3/2} + nP_{3/2} \to nS_{1/2} + (n+1)S_{1/2}$ in Rb Rydberg atoms can serve as an example [26]. In the absence of an electric field, they have small energy defects $\Delta = W(nS_{1/2}) + W([n+1]S_{1/2}) - 2 \times W(nP_{3/2})$, which depend on $n$. Here, $W(nL_J)$ denote the energies of Rydberg states in units of frequency. For $n \leq 38$ the energy defect can be made equal to zero due to the Stark effect in a dc electric field, while for higher states a combination of dc and radio-frequency fields should be used [27, 28]. At $\Delta = 0$ a resonant dipole-dipole interaction with an energy dependent on the distance as $R^{-3}$ arises between the atoms, while for large $\Delta$ there is a weaker van der Waals interaction proportional to $R^{-6}$. Thus, the character of interaction between Rydberg atoms can be changed significantly using an electric field and Förster resonances. For high states the exact Förster resonances cannot be tuned by an electric field (see Fig. 2 below). In this case, the interaction can also be described by the Förster resonance, but with a nonzero energy defect $\Delta$. It turns out to be more advantageous to use these states to observe the dipole blockade due to the large dipole moments.

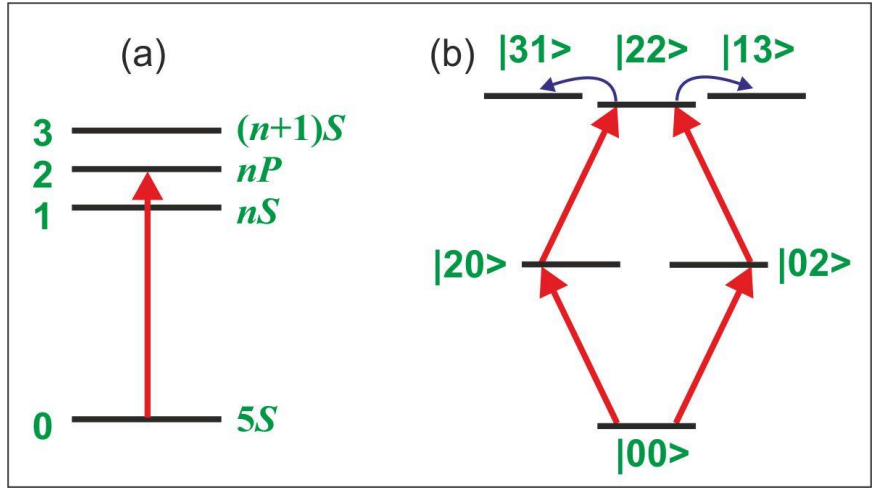

**Fig. 1.** (a) Scheme for the laser excitation of a single Rydberg Rb($nP$) atom and the states involved in the Förster resonance $nP_{3/2} + nP_{3/2} \to nS_{1/2} + (n+1)S_{1/2}$. (b) Scheme for the transitions between the collective states of two atoms. Resonant laser radiation causes the transitions between the states $|00\rangle \to (|02\rangle, |20\rangle) \to |22\rangle$, while the state $|22\rangle$ is coupled by the dipole-dipole interaction with the states $|31\rangle, |13\rangle$, which causes the transitions $|22\rangle \to |31\rangle, |13\rangle$. As a result of the interaction, the state $|22\rangle$ is shifted in energy, which leads to the dipole blockade effect.

To describe the dipole blockade effect under laser excitation of interacting Rydberg atoms, it is necessary to consider the transitions between various collective states of two atoms (Fig. 1b). Resonant laser radiation causes the transitions between the states $|00\rangle \to (|02\rangle, |20\rangle) \to |22\rangle$, while the state $|22\rangle$ is coupled by the dipole-dipole interaction operator with the states $|31\rangle, |13\rangle$, which causes the transitions $|22\rangle \to |31\rangle, |13\rangle$. As a result of the interaction, the state $|22\rangle$ is shifted in energy, which leads to the dipole blockade effect. The parameters of this problem are Rabi frequency $\Omega$ and the frequency detuning $\delta$ at the optical transition $0 \to 2$ in an individual atom, the matrix element of the resonant dipole-dipole interaction operator $V$ at the transitions $|22\rangle \to |31\rangle, |13\rangle$, and the Förster resonance energy defect $\Delta$.

For two Rydberg atoms in the initial state $nP_{3/2}(|M|=1/2)$ the matrix element of the dipole-dipole interaction operator is given by the expression

$$V = \frac{d_1 d_2}{4\pi\varepsilon_0}\left[\frac{1}{R^3} - \frac{3Z^2}{R^5}\right], \qquad (2)$$

where $d_1$ and $d_2$ are the $z$ components of the dipole moment matrix elements of the transitions $|nP_{3/2}(M_J = 1/2)\rangle \to |nS_{1/2}(M_J = 1/2)\rangle$ and $|nP_{3/2}(M_J = 1/2)\rangle \to |(n+1)S_{1/2}(M_J = 1/2)\rangle$, $Z$ is the $z$ component of the vector **R** connecting the two atoms (the $z$ axis is chosen along the control electric field), and $\varepsilon_0$ is the dielectric constant. Here, for simplicity, we take into account only the transitions without any change in the moment projection $M_J$, because otherwise it is necessary to also take into account the structure of the magnetic sublevels, which will complicate the problem dramatically. As was discussed in our paper [31], the energy shift $\delta W_{22}$ of the collective state $|22\rangle$ at the Förster resonance is described by the following approximate expression:

$$\delta W_{22} = \pm \left( \sqrt{\frac{\Delta^2}{4} + 2V^2} - \frac{|\Delta|}{2} \right). \tag{3}$$

Here, the sign is taken to be positive if the state $|22\rangle$ lies above the states $|31\rangle, |13\rangle$ and vice versa. At $\Delta = 0$ the interaction is a resonant dipole-dipole one, and the state $|22\rangle$ splits into two sublevels with energies $\pm\sqrt{2}V = C_3/R^3$, while at large $\Delta$ it becomes a van der Waals one with energy $\pm 2V^2/\Delta = C_6/R^6$, where $C_3$ and $C_6$ are the interaction constants [32].

The interacting atoms can be either spatially localized in separate optical dipole traps with a distance $R$ between them of a few micrometers [14, 15, 18–22] or be in a single laser excitation volume and have a random arrangement in it with some mean distance between atoms [8–13, 16, 17, 25–29]. Optical dipole traps are produced by tightly focusing nonresonant laser radiation [2]. The dipole blockade effect is usually observed for two atoms in neighboring traps with a distance of 5–10 μm between them. A single volume can be formed at the intersection of the focused laser beam exciting Rydberg states, as was done in our experiments [25–29]. The mean distance between atoms is then approximately half the excitation volume and is ~10 μm in our case. The dipole moments of Rydberg atoms are ~$n^2$ atomic units. From Eq. (2) we can then estimate the interaction energy and find that at $R$~10 μm it is necessary to excite the atoms to sufficiently high Rydberg states $n \geq 80$ to achieve dipole blockade in our experiments.

Note that, according to Eq. (2), the dipole-dipole interaction energy has a strong angular dependence and becomes zero at a certain angle between the quantization axis and vector **R**. Therefore, observing the dipole blockade for atoms with a random arrangement in the laser excitation volume requires a stronger interaction than that in the case of atoms in neighboring optical dipole traps for the probability of zero interaction energy to be low when averaging over the atomic positions. This, in turn, requires a closer arrangement of atoms or using higher Rydberg states, which is not always realizable in experiments. That is why only a partial dipole blockade was achieved in [8–13, 16, 17] for large atomic ensembles in a single volume, because the interaction energy was not enough to block the multiatom excitations in the entire volume.

To experimentally observe the dipole blockade effect when detecting atoms by the SFI method in a single volume, it is necessary to analyze the change in detection statistics and laser excitation spectra $S_N$ for a certain number $N$ of Rydberg atoms when passing from low ($n \leq 40$) to high ($n \geq 80$) Rydberg states by comparing them with the theoretical calculations for non-interacting atoms. Under a complete dipole blockade only one atom can be excited to a Rydberg state from the entire mesoscopic ensemble. Therefore, the dipole blockade should lead to a radical change in the multiatom spectra: only the single-atom spectrum $S_1$ should be observed, while all the remaining multiatom resonances should disappear. If they do not disappear completely, then this will suggest an incomplete dipole blockade, while a change in the ratio of the multiatom resonance amplitudes should allow the degree of completeness of the dipole blockade to be determined under specific experimental conditions.

Therefore, to perform our experiments, we chose the Rydberg $nP_{3/2}$ levels of Rb atoms with the principal quantum numbers $n$ = 39, 81, and 110. Their Stark maps calculated by the method from [33] are presented in Fig. 2 on the left. It can be seen on these maps that the Rydberg $S$, $P$, and $D$ states have large quantum defects and experience a quadratic Stark effect, in contrast to the hydrogen-like sets of levels with large orbital angular momenta that experience a linear Stark effect due to the energy degeneracy. The calculated Stark diagrams of the collective Rydberg states involved in the Förster resonances $nP_{3/2} + nP_{3/2} \rightarrow nS_{1/2} + (n+1)S_{1/2}$ for the sublevels with moment projection |$M_J$|=1/2 are presented in Fig. 2 on the right. The intersection of the collective states corresponds to the Förster resonance (resonant dipole-dipole interaction) and takes place only for the states with $n \leq 38$ due to the peculiar quantum defects and polarizabilities of the Rydberg levels of Rb atoms. We investigated such resonances previously [26–29]. The higher states in a dc electric field do not intersect and experience a van der Waals interaction, while the application of a dc field only increases the Förster resonance energy defect and weakens the interaction.

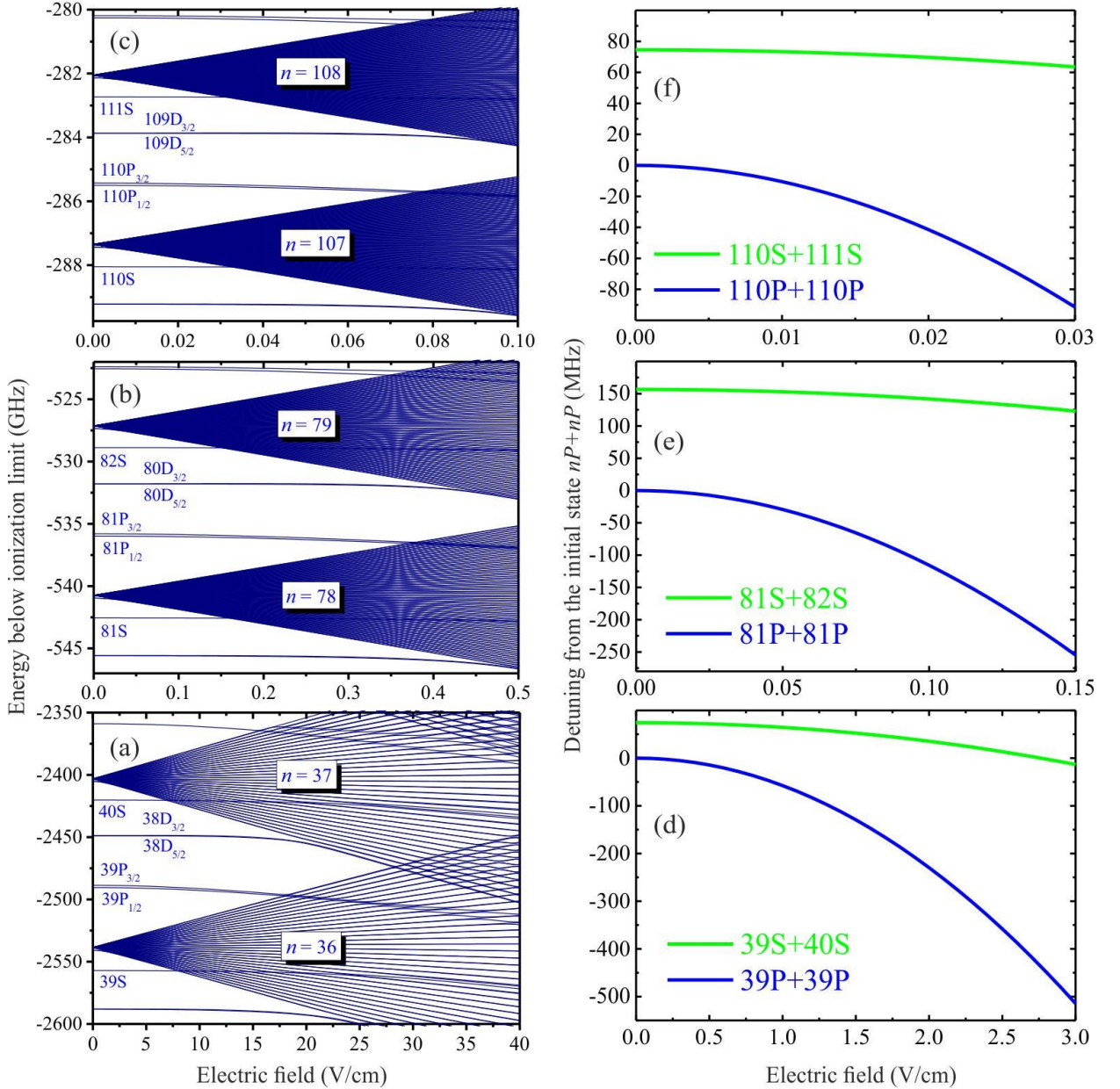

**Fig. 2.** Calculated Stark maps of the Rydberg levels of Rb atoms near the $nP_{3/2}$ states with $n$ = 39 (a), 81 (b), and 110 (c) for the sublevels with moment projection $|M_J|=1/2$. The Rydberg $S$, $P$, and $D$ states have large quantum defects and experience a quadratic Stark effect. The Stark diagrams of the collective Rydberg states involved in the Förster resonances $nP_{3/2} + nP_{3/2} \rightarrow nS_{1/2} + (n+1)S_{1/2}$ for the sublevels with moment projection $|M_J|=1/2$ are presented on panels (d–f). The intersection of the collective states corresponds to the Förster resonance (resonant dipole-dipole interaction) and takes place only for the states with $n \leq 38$. The higher states in a dc electric field do not intersect and experience a van der Waals interaction.

## 3. EXPERIMENTAL SETUP

The experiments are performed with cold $^{85}$Rb atoms captured into a magneto-optical trap (MOT), which is shown in Fig. 3a [3, 30]. The atoms are cooled by three orthogonal pairs of light waves with a wavelength of 780 nm. The cooling and repumping lasers are tuned to the closed transition $5S_{1/2}(F=3) \to 5P_{3/2}(F=4)$ and the transition $5S_{1/2}(F=2) \to 5P_{3/2}(F=3)$, respectively. A cloud of ~$10^6$ cold atoms 0.5–1 mm in size with a temperature of 100–200 μK is formed at the trap center.

The cold Rb atoms are excited to Rydberg states $nP$ ($n$ = 30–120) according to the three-step scheme $5S_{1/2} \to 5P_{3/2} \to 6S_{1/2} \to 37P_{3/2}$ (Fig. 3b). The first step, $55S_{1/2}(F=3) \to 5P_{3/2}(F=4)$, is excited by a Toptica DL PRO external-cavity diode laser with a wavelength of 780 nm operating in the continuous-wave mode with an output power up to 50 mW. The laser has a built-in Faraday isolator and an optical fiber output. The laser frequency is stabilized by the Pound-Drever-Hall technique based on the saturated-absorption resonance in a cell with vapors of Rb atoms. The measured laser line width is $\Gamma_1/(2\pi) \approx 0.3$ MHz. The output laser radiation is transmitted through an acousto-optic modulator (AOM), which produces pulses of arbitrary duration with the edges of ~100 ns and provides a "blue" detuning $\delta_1/(2\pi) = +80$ MHz from the exact atomic resonance lest the intermediate $5P_{3/2}$ level be populated.

In the second step, $5P_{3/2}(F=4) \to 6S_{1/2}(F=3)$, we use the radiation with a wavelength of 1367 nm from a Sacher TEC150 continuous-wave single-frequency external-cavity diode laser with a built-in Faraday isolator, an optical fiber output, and an output power up to 30 mW. The laser frequency is stabilized by the Pound-Drever-Hall technique based on the resonance of a highly stable optical Fabry-Perot interferometer by Stable Laser Systems ATF. The measured laser line width is $\Gamma_2/(2\pi) \approx 0.3$ MHz. The output laser radiation is transmitted through an electro-optic modulator with a modulation depth of 20 dB, which produces pulses of arbitrary duration with the edges of ~10 ns. The laser radiation frequency also has a blue detuning $\delta_2/(2\pi) = +82$ MHz from the exact atomic resonance lest the intermediate $6S_{1/2}$ level be populated.

In the third step, the Rydberg $nP$ states are excited from the state $6S_{1/2}(F=3)$ by the radiation from a Tekhnoscan TIS-SF-07 continuous-wave titanium-sapphire ring laser with an output power up to 500 mW. When the laser is tuned in the wavelength range 738–745 nm, the fine-structure sublevels $J = 1/2, 3/2$ of the Rydberg $nP$ states with principal quantum numbers $n$ = 30–120 can be excited selectively. The laser frequency is stabilized by the Pound-Drever-Hall technique based on the resonance of the same highly stable optical Fabry-Perot interferometer Stable Laser Systems ATF. The measured laser line width is $\Gamma_3/(2\pi) \approx$ 0.01 MHz. An AOM for operation in the pulsed mode with the edges of ~100 ns is installed at the laser output.

The radiations from the second- and third-step lasers are fed to MOT through single-mode optical fibers. At the exit from the optical fibers they are collimated and then focused on the cloud of cold atoms in the geometry of beams crossed at a right angle (Fig. 3a) with a waist diameter of ~10 μm for the 743-nm radiation and ~20 μm for the 1367-nm radiation. An effective excitation volume of Rydberg atoms with a size of ~20 μm, depending on the relative positions of the waists and the presence or absence of transition saturation, is formed in the region where the focused beams intersect. The first-step laser radiation (780 nm) is not focused, has a beam diameter of 1 mm, and is directed to the cloud of cold atoms at an angle of 45° toward the remaining beams. Rydberg atoms are excited by laser pulses with a repetition rate of 5 kHz.

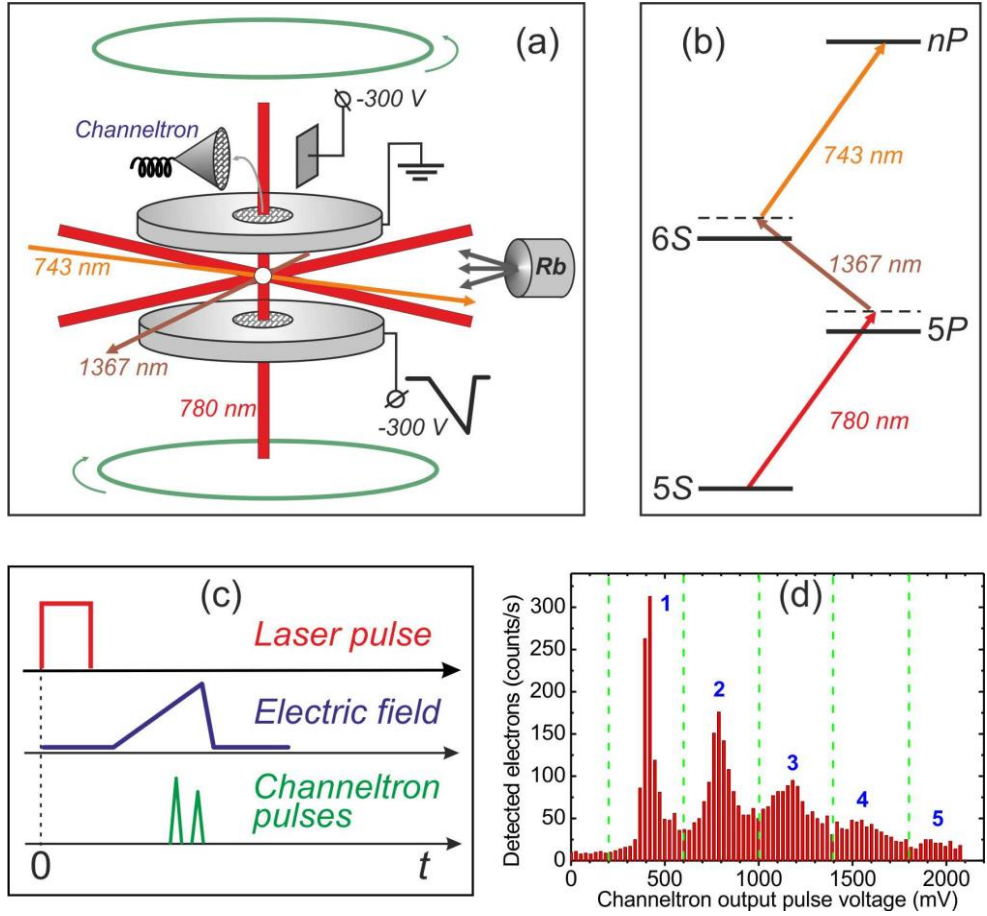

**Fig. 3.** (a) Scheme of the experiment with cold Rydberg $^{85}$Rb atoms in MOT. Rydberg atoms are excited in a small volume of the cloud of cold atoms and are detected by the SFI method. (b) Scheme of the coherent three-photon laser excitation $5S_{1/2} \to 5P_{3/2} \to 6S_{1/2} \to nP$ of Rydberg Rb atoms with a detuning from the intermediate resonances. (c) Time diagram of the laser and electric pulses. (d) Histogram of the output pulses of the secondary VEU-6 electron multiplier that detects the electrons produced by SFI. The individual peaks corresponding to $N = 1–5$ detected Rydberg atoms are observed.

The Rydberg atoms are excited in the space between two plates of stainless steel producing a uniform electric field (Fig. 3a). The electric field is used for Stark spectroscopy and the detection of Rydberg atoms by the SFI method. The atoms are detected with a repetition rate of 5 kHz when an ionizing electric field sweep pulse with a rise time of 2–3 μm is switched on. The electrons produced by ionization are accelerated by the electric field, fly through the metal grid of the upper plate, and are directed into the input mouth of the VEU-6 channel electron multiplier by the deflecting electrode. The pulsed signals from its output are processed by a high-speed ADC, a strobe integrator, and a computer. The number of electrons detected per laser pulse is determined by the number of Rydberg atoms in the excitation region and the total electron detection efficiency [25]. In our experiments the detection efficiency reaches 70% [27, 28].

A time diagram of the signals in the detection system is presented in Fig. 3c. An ionizing electric field sweep with a rise time of about 2 μs was switched on after each laser pulse exciting some of the cold atoms to a Rydberg $nP$ state. Depending on the state of a Rydberg atom, its ionization occurred at different instants of time after the laser pulse, according to Eq. (1). Then, the pulsed ionization signal was recorded at the VEU-6 output using a strobe pulse corresponding to the ionization of the $nP$ state in time. Figure 3d shows a histogram of the amplitudes of VEU-6 output pulses. Several peaks corresponding to different numbers of detected Rydberg atoms, $N = 1–5$, can be seen on this histogram. The integrated amplitude (area) of each of the peaks is described by a Poisson distribution and depends on the mean number of detected atoms per laser pulse. For Fig. 3d this quantity is 2.2 atoms per pulse; the single-atom and two-atom excitation probabilities are approximately equal, but the single-atom peak is narrower and higher.

After each laser pulse, the data acquisition system measured the VEU-6 output pulse amplitude, then determined the number of detected atoms from the premeasured histogram (Fig. 3d), and sorted the signals by the number of atoms $N$ and calculated the three-photon laser excitation probability of a Rydberg state after the accumulation of data in $10^3$-$10^4$ laser pulses. The number of atoms was determined in accordance with the interval of VEU-6 output pulse voltages into which a specific measured signal fell: for example, 200-600 mV for one atom, 600-1000 mV for two atoms, 1000-1400 mV for three atoms, etc. The corresponding signal thresholds are indicated in Fig. 3d by the vertical dashed lines.

Our experiments on three-photon excitation spectroscopy were performed in MOT switched off in advance for a short time. For this purpose, acousto-optic modulators were installed on all the cooling laser beams, which switched them off for 20 μs, and were switched on again after the measurement. The MOT gradient magnetic field was not switched off during our measurements, but its influence was minimized by adjusting the position of the excitation volume to the point of zero magnetic field, which was controlled by the absence of Zeeman splitting of the microwave transition $37P_{3/2} \to 37S_{1/2}$ at 80 GHz by our method from Ref. [34]. This allowed us to have a high laser pulse repetition frequency (5 kHz) and to trace the change in the signals from Rydberg atoms in real time on the oscilloscope screen and in the computer-based data acquisition system.

## 4. THREE-PHOTON LASER EXCITATION SPECTRA OF MESOSCOPIC ENSEMBLES OF RYDBERG ATOMS

The three-photon laser excitation spectra for the Rydberg states $39P_{3/2}$, $81P_{3/2}$, and $110P_{3/2}$ were recorded by slowly scanning the radiation frequency of the third-step laser (titanium-sapphire laser). For this purpose, we tuned the frequency of the digital synthesizer controlling the electro-optic modulator in the laser frequency stabilization system based on the highly stable reference Fabry-Perot interferometer (one of the side frequencies of the laser radiation transmitted through the modulator was locked to the interferometer, which made it possible to tune the frequency of the main radiation in the regime of frequency stabilization). Our records are presented in Fig. 4. Column $S$ is the signal corresponding to the mean number of detected Rydberg atoms per laser pulse. Columns $S_1$-$S_5$ are the excitation spectra for mesoscopic ensembles with a certain number of Rydberg atoms, $N = 1–5$. Their sum gives the total measured signal $S$. The dipole blockade for the high Rydberg states should reduce the resonance amplitudes for $N = 2–5$ and increase the resonance amplitude for $N = 1$, which requires their comparison with the theoretical calculations.

Subsequently, we analyzed the laser excitation spectra of the Rydberg states $39P_{3/2}$, $81P_{3/2}$, and $110P_{3/2}$ for the presence of dipole blockade signatures in them. The low Rydberg state $39P_{3/2}$ has a Förster resonance energy defect of 74.3 MHz in zero electric field (see Fig. 2d) and the calculated matrix element of the dipole-dipole interaction operator $V(39P) \approx 0.46$ MHz for the mean distance between atoms $R = 10$ μm. Hence from Eq. (3) we obtain an estimate for the energy shift of the collective state $|22\rangle$: $\delta W_{22} \approx 6$ kHz, which is negligible compared to the width of the laser excitation spectrum for this state, $\delta \nu \approx 2.3$ MHz (Fig. 4a). Therefore, no dipole blockade effect is expected for this state, while its multiatom spectra can be used as reference ones for non-interacting atoms and can be compared with the spectra for the high states.

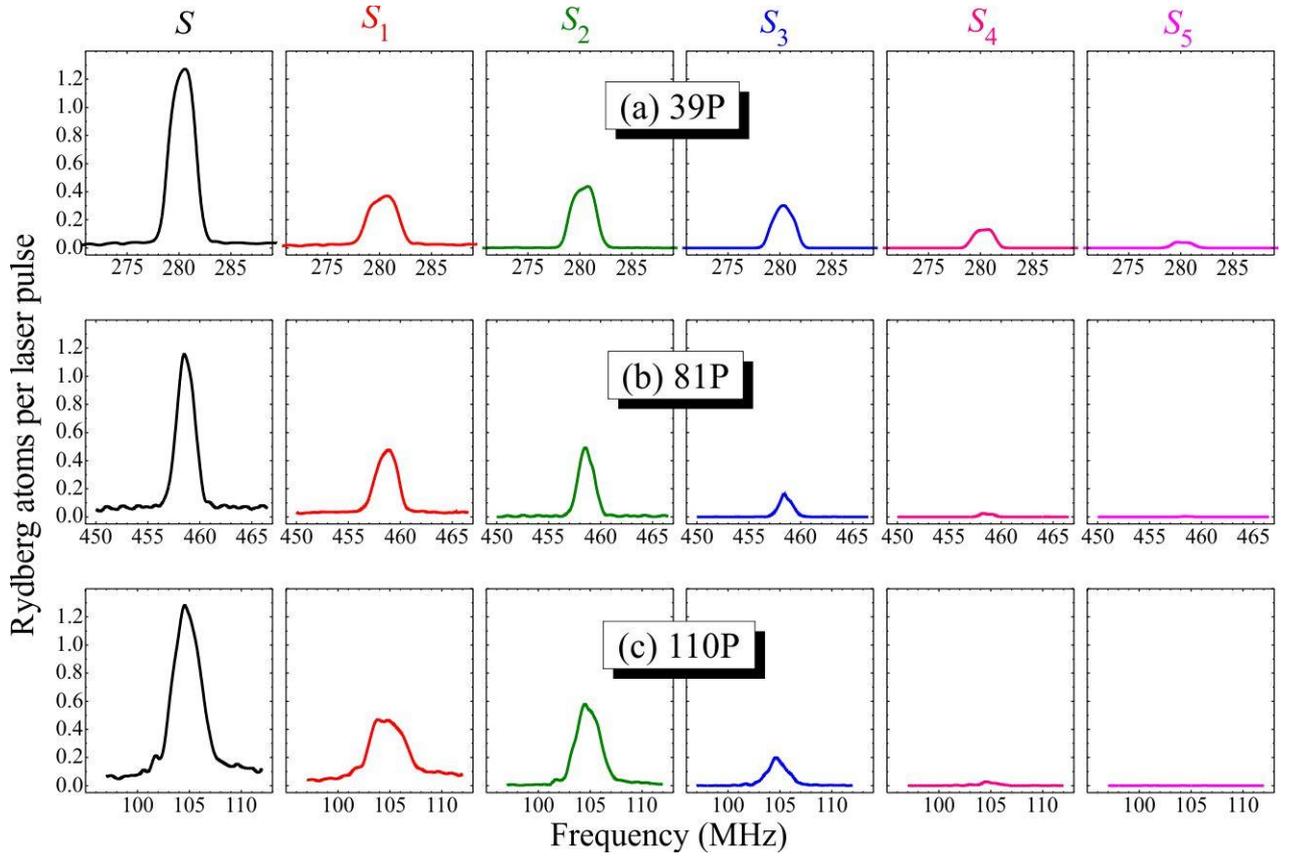

**Fig. 4.** Experimental records of the three-photon laser excitation spectra for the Rydberg states $39P_{3/2}$ (a), $81P_{3/2}$ (b), and $110P_{3/2}$ (c). Column $S$ is the signal corresponding to the mean number of detected Rydberg atoms per laser pulse. Columns $S_1$-$S_5$ are the excitation spectra for mesoscopic ensembles with a certain number of Rydberg atoms, $N$ = 1–5. Their sum gives the total measured signal $S$. The dipole blockade for the high Rydberg states should reduce the resonance amplitudes for $N$ = 2–5 and increase the resonance amplitude for $N$ = 1, which requires their comparison with the theoretical calculations.

The high Rydberg state $81P_{3/2}$ has a Förster resonance energy defect of 156.6 MHz (see Fig. 2e) and the calculated matrix element of the dipole-dipole interaction operator $V(81P) \approx$ 10 MHz for the mean distance between atoms $R$ = 10 μm. Hence from Eq. (3) we obtain an estimate for the energy shift of the collective state $|22\rangle$: $\delta W_{22} \approx 1.3$ MHz, which is already comparable to the width of the laser excitation spectrum for this state, $\delta\nu \approx 1.7$ MHz (Fig. 4b). Therefore, for this state one might expect a partial dipole blockade.

Finally, the even higher Rydberg state $110P_{3/2}$ has a Förster resonance energy defect of 74.6 MHz (see Fig. 2f) and the calculated matrix element of the dipole-dipole interaction operator $V(110P) \approx 35$ MHz for the mean distance between atoms $R$ = 10 μm. Hence from Eq. (3) we obtain an estimate for the energy shift of the collective state $|22\rangle$: $\delta W_{22} \approx 25$ MHz, which exceeds considerably the width of the laser excitation spectrum for this state, $\delta\nu \approx 2.8$ MHz (Fig. 4c). However, this state has a very high polarizability in a dc electric field and intersects with the hydrogen-like set of Rydberg states $n$ = 107 already in a 80 mV cm$^{-1}$ field, while the neighboring $S$ states involved in the Förster resonance intersect with different levels already in a 30 mV cm$^{-1}$ field (see Fig. 2c). Meanwhile, there is always an uncontrollable parasitic electric field in the experiments (for example, from the Rb atoms that settled on the electric plates of the detection system and from the VEU-6 multiplier with a supply voltage of 3–4 kV). The previously conducted experiments on microwave spectroscopy have shown that there is a parasitic electric field of ~50 mV cm$^{-1}$ in our detection system. It can increase the energy defect and even destroy the Förster resonance, which, in turn, weakens the dipole blockade. Hence, only a partial dipole blockade was also expected for the state $110P_{3/2}$, but stronger than that for the state $81P_{3/2}$.

Yet another problem is the decrease in the probability of detecting atoms in high Rydberg states by the SFI method. According to Eq. (1), the critical electric fields for SFI are 8.5 V cm$^{-1}$ for the state $81P_{3/2}$ and 2.4 V cm$^{-1}$ for the state $110P_{3/2}$, while for the low state $39P_{3/2}$ this field is 183 V cm$^{-1}$. Since the electron detached from an atom reaches VEU-6 in a time of 10–100 ns, which is much shorter than the rise time of the ionizing pulse, the electron energy is virtually equal to the energy gained in the ionizing field. The detection probability measured by us previously for the low states with $n = 36$ and 37 from the ratios of the single-atom and two-atom signals at Förster resonances was 70% [27, 28]. We did not make such measurements for high Rydberg states, because there are no Förster resonances for them in a dc electric field (see Fig. 2). However, it is known from the published data that the maximum electron detection efficiency of VEU-6 corresponds to energies of 100–300 eV; it decreases approximately by a factor of 1.5 and 2.3 at electron energies of 10 and 2 eV, respectively [35]. In addition, since the MOT gradient magnetic field in our experiment is not switched off, the low-energy electrons fly to VEU-6 in a more complex trajectory, which can lead to a further decrease in the detection efficiency. Therefore, for high Rydberg states, particularly for the state $110P_{3/2}$, one would expect that the influence of the dipole blockade effect on the multiatom excitation spectra can be weakened by the lower probability of detecting Rydberg atoms by the SFI method.

The spectra $S_N$ for the state $39P_{3/2}$ (see Fig. 4a) were taken as reference multiatom spectra for non-interacting atoms (without any dipole blockade). It can be seen even from a direct comparison of these spectra with those for the states $81P_{3/2}$ (see Fig. 4b) and $110P_{3/2}$ (see Fig. 4c) that the multiatom resonance amplitudes $S_3$-$S_5$ for the high states are significantly reduced compared to the resonance amplitude for the $39P_{3/2}$ state. This is the first signature of a partial dipole blockade for the high states.

## 5. DIPOLE BLOCKADE IN THE THREE-PHOTON LASER EXCITATION SPECTRA AND STATISTICS FOR MESOSCOPIC ENSEMBLES OF RYDBERG ATOMS

To obtain quantitative information about the dipole blockade, we should consider the laser excitation and detection statistics of non-interacting atoms and compare it with the statistics measured for high states. For this purpose, we will use the theory of laser excitation and detection statistics of non-interacting atoms developed by us previously [30]. Let we have a mesoscopic ensemble of $N_0$ ground-state Rb atoms in the excitation volume before the beginning of a laser pulse. There is a nonzero probability of exciting each of the atoms to a Rydberg state in the laser pulse time, $0 \leq p \leq 1$, with the probability $p$ depending on the three-photon detuning and being the excitation spectrum of a single atom. The mean number of Rydberg atoms excited per laser pulse is then

$$\bar{n} = p N_0. \qquad (4)$$

The statistics of Rydberg atoms excited per laser pulse depends on $p$. Under weak excitation ($p \ll 1$) we can apply the Poisson distribution $P_N^{weak}$ for the probability to find $N$ Rydberg atoms after an individual laser pulse:

$$P_N^{weak} = \frac{(\bar{n})^N}{N!} e^{-\bar{n}}. \tag{5}$$

However, in the general case, including strong coherent excitation with Rabi oscillations, a more complex normal distribution should be applied:

$$P_N^{strong} = p^N (1-p)^{N_0-N} \frac{N_0!}{N!(N_0-N)!}, \tag{6}$$

which is valid for any $p$ and $N_0$. Such a statistical distribution would be observed for an ideal detector of Rydberg atoms with the detection probability $T = 1$.

For real detectors the detection probability is always less than 1. Taking the convolution of the probability of exciting and detecting a certain number of atoms, it can be shown that, given the finite detection probability, we will have the following distributions for the probability to detect $N$ Rydberg atoms:

$$\bar{P}_N^{weak} = \frac{(\bar{n}T)^N}{N!} e^{-\bar{n}T}, \tag{7}$$

$$\bar{P}_N^{strong} = (pT)^N (1-pT)^{N_0-N} \frac{N_0!}{N!(N_0-N)!}. \tag{8}$$

Thus, the mean number of Rydberg atoms detected per laser pulse decreases to $\bar{n}T$. This value is measured experimentally when averaged over a large number of laser pulses. For example, the amplitudes of the peaks on the histogram in Fig. 3d are proportional to $\bar{P}_N^{weak}$. Therefore, the ratio between the integrated single-atom and two-atom peaks is $\bar{P}_2^{weak}/\bar{P}_1^{weak} = \bar{n}T/2$, and our measurement for this histogram gives $\bar{n}T \approx 2{,}2$.

Under strong laser excitation the total signal measured as the mean number of atoms detected per laser pulse is given by the expression

$$S = pN_0 T = \sum_{N=1}^{N_0} N \bar{P}_N^{strong} = \sum_{N=1}^{N_0} S_N, \tag{9}$$

i.e., the total signal $S$ in Fig. 4a for the non-interacting state $39P_{3/2}$ is actually averaged over a different number $N$ of detected atoms. The solution of the inverse problem gives the following formula for the multiatom spectra in Fig. 4a:

$$S_N = N \left(\frac{S}{N_0}\right)^N \left(1 - \frac{S}{N_0}\right)^{N_0-N} \frac{N_0!}{N!(N_0-N)!} \tag{10}$$

Thus, for a known number of ground-state atoms $N_0$ the multiatom spectra $S_N$ for the non-interacting state $39P_{3/2}$ in Fig. 4a should be uniquely determined by the total spectrum $S$, which is simply the mean number of Rydberg atoms detected per laser pulse.

According to the premeasured data, the cloud of cold Rb atoms in our magneto-optical trap 0.5–0.7 mm in diameter contains $\sim 10^6$ ground-state atoms. Then, $N_0 \approx 50$ atoms will be contained in the laser excitation volume of Rydberg atoms with a size of $\sim 20$ μm. At the same time, Eq. (10) ceases to be sensitive to the exact value of $N_0$ already for 10 atoms. Therefore, no measurement of the exact number of atoms is required for its comparison with the experiment.

Figure 5 compares the experimental records of the multiatom three-photon laser excitation spectra $S_N$ for the Rydberg states $39P_{3/2}$, $81P_{3/2}$, and $110P_{3/2}$ with the spectra calculated from Eq. (10) for non-interacting atoms at $N_0 \approx 50$. It can be seen from this figure that for the non-interacting state $39P_{3/2}$ there is almost complete agreement between the experiment and theory in multiatom resonance amplitudes and shapes irrespective of the number $N$ of detected atoms. At the same time, for the high Rydberg states $81P_{3/2}$ and $110P_{3/2}$ the expected partial dipole blockade reduces the resonance amplitudes for $N = 3$–5 and increases them for $N = 1, 2$.

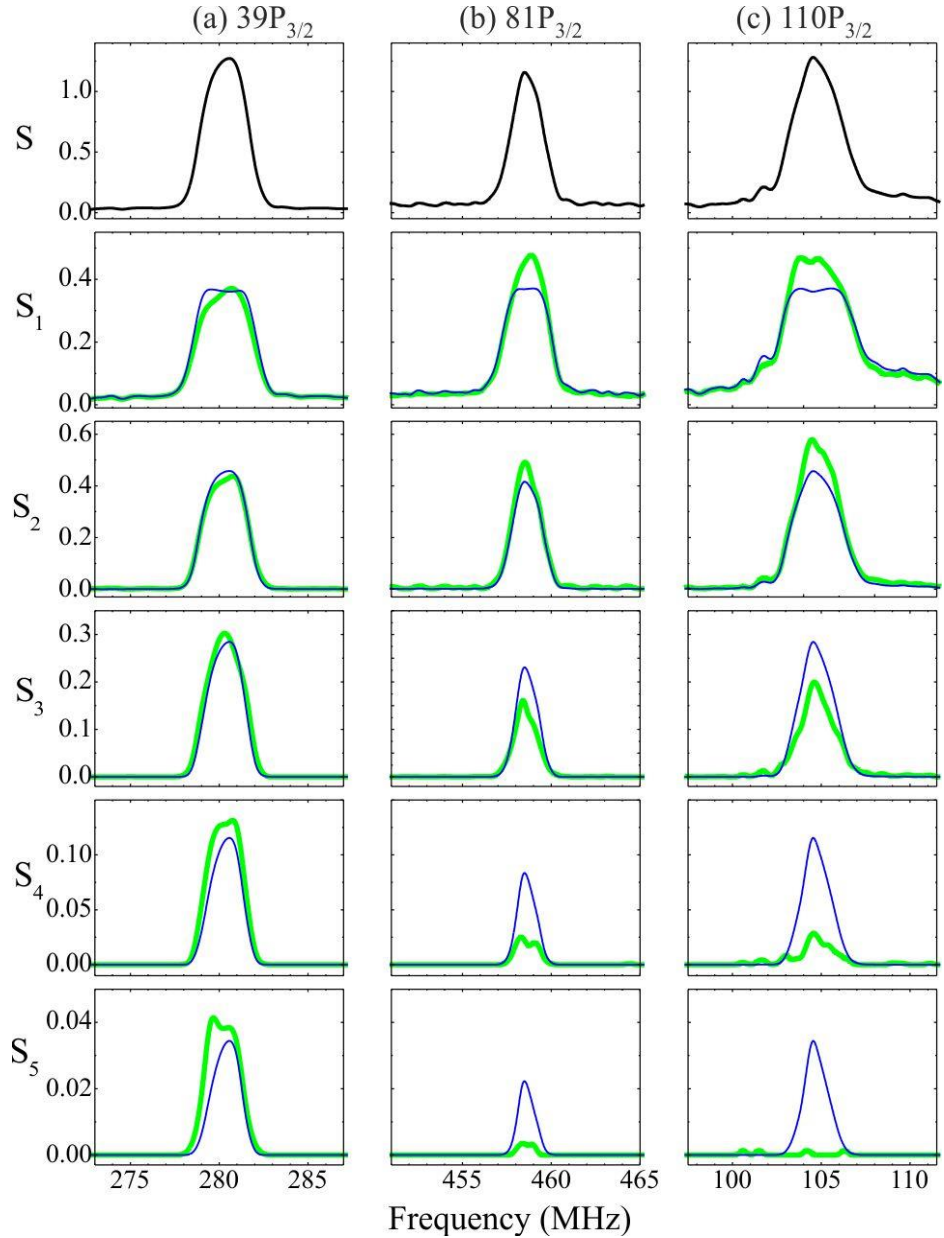

**Fig. 5.** Comparison of the experimental records of the multiatom three-photon laser excitation spectra $S_N$ for the Rydberg states $39P_{3/2}$ (a), $81P_{3/2}$ (b), and $110P_{3/2}$ (c) (thick green curves) with the spectra calculated from Eq. (10) for non-interacting atoms at $N_0 \approx 50$ (thin blue curves). The partial dipole blockade for the high Rydberg states $81P_{3/2}$ and $110P_{3/2}$ reduces the resonance amplitudes for $N = 3$–5 and increases them for $N = 1, 2$.

For our quantitative measurements Fig. 6 compares the experimental amplitudes of the multiatom three-photon laser excitation spectra $S_N$ for the Rydberg states $39P_{3/2}$, $81P_{3/2}$, and $110P_{3/2}$ with the amplitudes of the spectra calculated from Eq. (10) for non-interacting atoms at $N_0 \approx 50$. It can also be seen from this figure that for the non-interacting state $39P_{3/2}$ there is almost complete agreement between the experiment and theory in multiatom resonance amplitudes irrespective of the number $N$ of detected atoms.

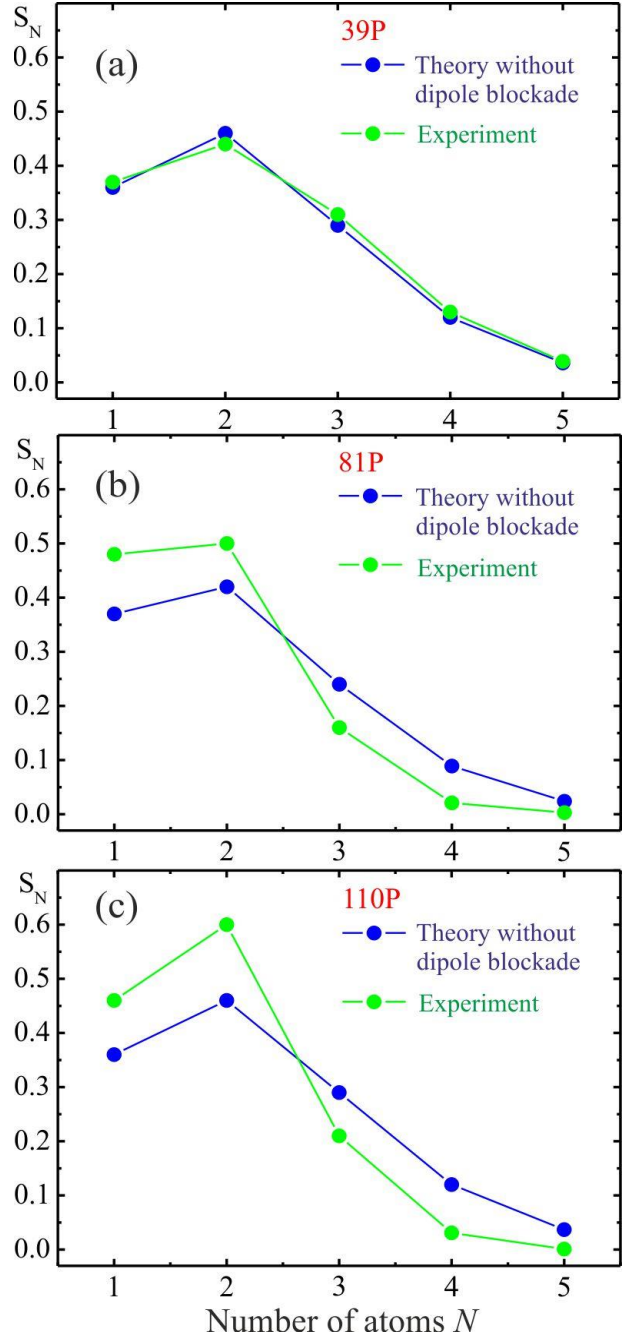

**Fig. 6.** Comparison of the experimental amplitudes of the multiatom three-photon laser excitation spectra $S_N$ for the Rydberg states $39P_{3/2}$ (a), $81P_{3/2}$ (b), and $110P_{3/2}$ (c) with the amplitudes of the spectra calculated from Eq. (10) for non-interacting atoms at $N_0 \approx 50$. The measurement error roughly corresponds to the sizes of the circles representing the experimental data points. The partial dipole blockade for the high Rydberg states $81P_{3/2}$ and $110P_{3/2}$ reduces the resonance amplitudes for $N = 3$–5 and increases them for $N = 1, 2$.

For the high Rydberg state $81P_{3/2}$ the partial dipole blockade increases the single-atom signal from the theoretical value for non-interacting atoms, $S_1^T = 0.37$, to the experimental value for interacting atoms, $S_1^E = 0.48$, and increases the two-atom signal from $S_2^T = 0.42$ to $S_2^E = 0.5$ (Fig. 6b). At the same time, the three-atom signal decreases from $S_3^T = 0.24$ to $S_3^E = 0.16$, the four-atom one decreases from $S_4^T = 0.089$ to $S_4^E = 0.021$, and the five-atom one decreases from $S_5^T = 0.024$ to $S_5^E = 0.003$. If we renormalize these values to identical single-atom signals for a proper comparison, then all theoretical values should be multiplied by the normalization factor $S_1^E / S_1^T = 1.3$. Finally, we then find that the partial dipole blockade does not change the population of the two-atom collective state, but it reduces the population of the three-atom, four-atom, and five-atom states by 50 ± 5%, 81 ± 7%, and 90 ± 8%, respectively.

Similar results are obtained from our analysis of Fig. 6c for the higher Rydberg state $110P_{3/2}$ as well. The partial dipole blockade for it does not change the population of the two-atom collective state either, but it reduces the population of the three-atom, four-atom, and five-atom states by 50 ± 5%, 83 ± 7%, and 97 ± 3%, respectively. Thus, for this state the dipole blockade has an even greater influence on the four-atom and five-atom signals, with the five-atom signals being suppressed almost completely. Complete dipole blockade for this state is not observed for the reasons discussed above.

Yet another confirmation of the change in the laser excitation statistics for the high Rydberg states $81P_{3/2}$ and $110P_{3/2}$ due to the partial dipole blockade compared to the statistics for the non-interacting state $39P_{3/2}$ is a comparison of the histograms of VEU-6 multiplier output pulses for different numbers $N$ of detected Rydberg atoms when tuning to the centers of the lines of the laser transitions to these states (Fig. 7). The partial dipole blockade for the high states $81P_{3/2}$ and $110P_{3/2}$ reduces the signal amplitudes for $N = 3$–5 and increases them for $N = 1, 2$. One of the methods for the detection of a partial dipole blockade in such histograms is to measure the Mandel parameter $Q$, which should be zero for purely Poissonian statistics (non-interacting atoms) and less than zero under dipole blockade conditions, as was demonstrated in Refs. [10, 16]. However, we did not make such measurements, because the change in the ratio of the amplitudes in Fig. 6 already clearly shows sub-Poissonian detection statistics and a partial dipole blockade for the high Rydberg states.

## 6. CONCLUSIONS

In this paper we presented our experimental results on the observation of the dipole blockade for mesoscopic ensembles of $N = 1$–5 atoms in a single trap detected by the SFI method. We investigated the multiatom spectra of the three-photon laser excitation $5S_{1/2} \to 5P_{3/2} \to 6S_{1/2} \to nP_{3/2}$ of cold Rydberg Rb atoms localized in a small excitation volume (~20 μm in size) in a magneto-optical trap. No signatures of the dipole blockade were detected for the low Rydberg state $39P_{3/2}$ at a mean distance between atoms of ~10 μm, while for the high states $81P_{3/2}$ and $110P_{3/2}$ a significant decrease in the resonance amplitudes was observed for $N = 3$–5, suggesting that the regime of a partial dipole blockade was reached. At the same time, we failed to observe a complete dipole blockade, when only the resonances with $N = 1$ remain. This is most likely due to the presence of parasitic electric fields reducing the interaction energy of Rydberg atoms, the decrease in the probability of detecting high states by the SFI method, and the strong angular dependence of the interaction energy of Rydberg atoms in a single interaction volume.

Based on our results, we may conclude that the optical method is more preferable when the dipole blockade is detected for high Rydberg states, despite its considerably lower speed. Our numerical calculations in [31] also showed that to achieve a complete dipole blockade in our experiment, it is necessary either to reduce the available laser excitation volume by a factor of 2–3 (to a size less than 10 μm) or to perform experiments for single atoms in neighboring optical dipole traps. The latter will eliminate the angular dependence of the interaction energy of Rydberg atoms, which weakens the dipole blockade when averaging over the single interaction volume, as in our experiment.

This work was supported by the Russian Foundation for Basic Research (project no. 19-52-15010 with regard to the theoretical analysis of Förster resonances and project no. 17-02-00987 with regard to the applications in quantum informatics), the Russian Science Foundation (project no. 18-12-00313 with regard to the dipole blockade theory), the Advanced Research Foundation (with regard to the experiment and analysis of its results), and the Novosibirsk State University.

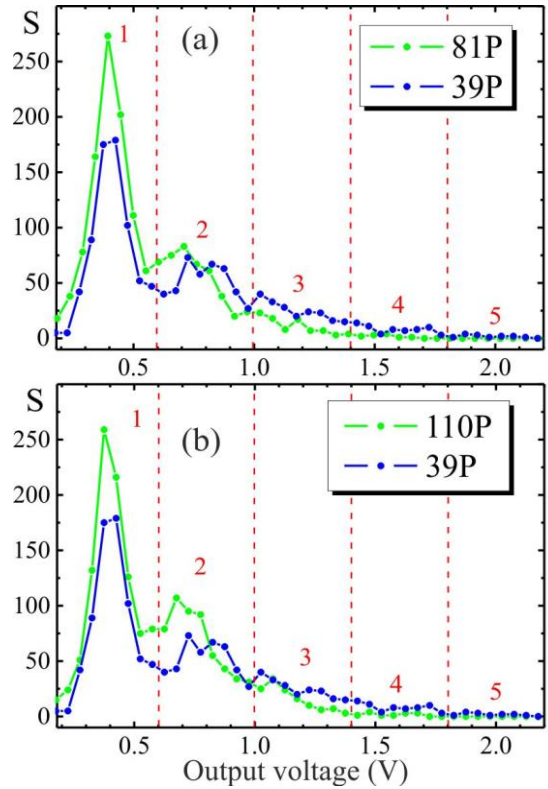

**Fig. 7.** Comparison of the histograms of VEU-6 multiplier output pulses for different numbers $N$ of detected Rydberg atoms when tuning to the centers of the lines of the laser transitions to the non-interacting state $39P_{3/2}$ and the interacting states $81P_{3/2}$ (a) and $110P_{3/2}$ (b). The partial dipole blockade for the high Rydberg states $81P_{3/2}$ and $110P_{3/2}$ reduces the signal amplitudes for $N = 3$–$5$ and increases them for $N = 1, 2$.

**REFERENCES**

[1] T.F. Gallagher, *Rydberg atoms* (Cambridge: Cambridge University Press, 1994).
[2] M. Saffman, T. G. Walker, K. Mølmer, Rev. Mod. Phys. **82**, 2313 (2010).
[3] I.I. Ryabtsev, I.I. Beterov, D.B. Tretyakov, V.M. Entin, E.A. Yakshina, Phys. Usp. **59**, 196 (2016).
[4] M. Saffman, J. Phys. B **49**, 202001 (2016).
[5] D. Jaksch, J.I. Cirac, P. Zoller, S.L. Rolston, R. Cote, M.D. Lukin, Phys. Rev. Lett., **85**, 2208 (2000).
[6] M.D. Lukin, M. Fleischhauer, R. Cote, L.M. Duan, D. Jaksch, J.I. Cirac, P. Zoller, Phys. Rev. Lett., **87**, 037901 (2001).

[7] D. Comparat, P. Pillet, J. Opt. Soc. Am. B, **27**, A208 (2010).
[8] D. Tong, S.M. Farooqi, J. Stanojevic, S. Krishnan, Y.P. Zhang, R. Côté, E.E. Eyler, P.L. Gould, Phys. Rev. Lett., **93**, 063001 (2004).
[9] K. Singer, M. Reetz-Lamour, T. Amthor, L.G. Marcassa, M. Weidemüller, Phys. Rev. Lett., **93**, 163001 (2004).
[10] T. Cubel Liebisch, A. Reinhard, P.R. Berman, G. Raithel, Phys. Rev. Lett., **95**, 253002 (2005).
[11] T. Vogt, M. Viteau, J. Zhao, A. Chotia, D. Comparat, P. Pillet, Phys. Rev. Lett., **97**, 083003 (2006).
[12] T. Vogt, M. Viteau, A. Chotia, J. Zhao, D. Comparat, P. Pillet, Phys. Rev. Lett., **99**, 073002 (2007).
[13] R. Heidemann, U. Raitzsch, V. Bendkowsky, B. Butscher, R. Löw, L. Santos, T. Pfau, Phys. Rev. Lett., **99**, 163601 (2007).
[14] T. Wilk, A. Gaetan, C. Evellin, J. Wolters, Y. Miroshnichenko, P. Grangier, A. Browayes, Phys. Rev. Lett., **104**, 010502 (2010).
[15] L. Isenhower, E. Urban, X.L. Zhang, A.T. Gill, T. Henage, T.A. Johnson, T.G. Walker, M. Saffman, Phys. Rev. Lett., **104**, 010503 (2010).
[16] M. Viteau, P. Huillery, M.G. Bason, N. Malossi, D. Ciampini, O. Morsch, E. Arimondo, D. Comparat, P. Pillet, Phys. Rev. Lett., **109**, 053002 (2012).
[17] A. Schwarzkopf, R.E. Sapiro, G. Raithel, Phys. Rev. Lett., **107**, 103001 (2011).
[18] A.M. Hankin, Y.-Y. Jau, L.P. Parazzoli, C.W. Chou, D.J. Armstrong, A.J. Landahl, G.W. Biedermann, Phys. Rev. A, **89**, 033416 (2014).
[19] D. Barredo, S. Ravets, H. Labuhn, L. Béguin, A. Vernier, F. Nogrette, T. Lahaye, A. Browaeys, Phys. Rev. Lett., **112**, 183002 (2014).
[20] M. Ebert, M. Kwon, T.G. Walker, M. Saffman, Phys. Rev. Lett., **115**, 093601 (2015).
[21] Y.-Y. Jau, A.M. Hankin, T. Keating, I.H. Deutsch, G.W. Biedermann, Nature Physics **12**, 71 (2016).
[22] Y. Zeng, P. Xu, X. He, Y. Liu, M. Liu, J. Wang, D.J. Papoular, G.V. Shlyapnikov, M. Zhan, Phys. Rev. Lett., **119**, 160502 (2017).
[23] Y.O. Dudin, L. Li, F. Bariani, A. Kuzmich, Nature Physics **8**, 790 (2012).
[24] Y.O. Dudin, A. Kuzmich, Science **336**, 887 (2012).
[25] I.I. Ryabtsev, D.B. Tretyakov, I.I. Beterov, and V.M. Entin, Phys. Rev. A **76**, 012722 (2007); *Erratum:* Phys. Rev. A, **76**, 049902(E) (2007).
[26] I.I. Ryabtsev, D.B. Tretyakov, I.I. Beterov, and V.M. Entin, Phys. Rev. Lett., **104**, 073003 (2010).
[27] D.B. Tretyakov, V.M. Entin, E.A. Yakshina, I.I. Beterov, C. Andreeva, I.I. Ryabtsev, Phys. Rev. A, **90**, 041403(R) (2014).
[28] E.A. Yakshina, D.B. Tretyakov, I.I. Beterov, V.M. Entin, C. Andreeva, A. Cinins, A. Markovski, Z. Iftikhar, A. Ekers, I.I. Ryabtsev, Phys. Rev. A **94**, 043417 (2016).
[29] D.B. Tretyakov, I.I. Beterov, E.A. Yakshina, V.M. Entin, I.I. Ryabtsev, P. Cheinet, P. Pillet, Phys. Rev. Lett., **119**, 173402 (2017).
[30] E.A. Yakshina, D.B. Tretyakov, V.M. Entin, I.I. Beterov, I.I. Ryabtsev, Quantum Electronics, **48** (10), 886 (2018).
[31] I.I. Ryabtsev, I.I. Beterov, D.B. Tretyakov, E.A. Yakshina, V.M. Entin, Quantum Electronics, **49** (5), 455 (2019).
[32] A.A. Kamenski, N.L. Manakov, S.N. Mokhnenko, V.D. Ovsiannikov, Phys. Rev. A, **96**, 032716 (2017).
[33] L. Zimmerman, M.G. Littman, M.M. Kash, D. Kleppner, Phys. Rev. A, **20**, 2251 (1979).
[34] D.B. Tretyakov, I.I. Beterov, V.M. Entin, I.I. Ryabtsev, P.L. Chapovsky, J. Exp. Theor. Phys., **108**, 374 (2009).
[35] V.A. Rykov, P.P. D'yachenko, and A.A. Koshelev, Sov. At. Energy **63**, 539 (1987).